\documentclass[aps,twocolumn,prl,showpacs]{revtex4}

\usepackage{graphicx}
\usepackage{amsmath}
\usepackage{epsfig}
\usepackage{dcolumn}
\usepackage{bm}

\begin{document}

\title{One-Particle Anomalous Excitations of Gutzwiller-Projected BCS Superconductors and Bogoliubov Quasi-Particle Characteristics} 

\author{Seiji Yunoki}

\affiliation{
Istituto Nazionale per la Fisica della Materia and International 
School for Advanced Studies, via Beirut 4, 34014 Trieste, Italy,\\
Department of Physics and Astronomy, The University of Tennessee, 
Knoxville, Tennessee 37996, USA, and \\
Materials Science and Technology Division, Oak Ridge National Laboratory, 
Oak Ridge, Tennessee 32831, USA.
}

\date{\today}

\begin{abstract}

Low-lying one-particle anomalous excitations are studied for 
Gutzwiller-projected strongly correlated BCS states. It is found that 
the one-particle anomalous excitations are highly coherent, and the 
numerically calculated spectrum can be reproduced quantitatively by a 
renormalized BCS theory, thus strongly indicating that the nature of 
low-lying excitations described by the projected BCS states is essentially 
understood within a renormalized 
Bogoliubov quasi-particle picture. This finding resembles the well-known fact 
that a Gutzwiller-projected Fermi gas is a Fermi liquid. The present results 
are consistent with numerically exact calculations of the two-dimensional 
$t$-$J$ model as well as recent photoemission experiments on 
high-$T_{\rm C}$ cuprate superconductors.  
\end{abstract}

\pacs{74.20.Mn, 71.10.-w, 74.72.-h}

\maketitle


Since their discovery in 1986~\cite{bed}, high-$T_{\rm C}$ cuprate 
superconductors and related strongly correlated electronic 
systems have 
become one of the largest and most important fields in current condensed 
matter physics~\cite{rvb}. In spite of enormous theoretical and 
experimental efforts devoted since their discovery, understanding the 
mechanism of the superconductivity and the unusual normal state properties 
of the high-$T_{\rm C}$ cuprate superconductors remains unresolved and 
there is still extensive debate. 
This is mainly because of the strong-correlation nature of the problems, 
which is widely believed to be a key ingredient. 
Immediately after the discovery, Anderson~\cite{pwa} proposed 
a Gutzwiller-projected BCS state to incorporate strong-correlation 
effects in the superconducting state. While the ground 
state properties of projected BCS states have been studied 
extensively~\cite{yokoyama}, it is only very recently that the 
low-energy excitations of projected BCS states have been explored 
by numerically exact variational Monte Carlo 
techniques~\cite{rand,yunoki,fukushima,li}, which in fact have found many 
qualitative as well as very often quantitative similarities to the main 
features observed experimentally in high-$T_{\rm C}$ cuprate 
superconductors. It is therefore 
very important and timely to understand the nature of low-lying 
excitations described by Gutzwiller-projected BCS states because of their 
great relevance to the high-$T_{\rm C}$ cuprate superconductors. 
Furthermore, a Gutzwiller projected single-particle state is one of the most 
widely used correlated many-body wave functions in a variety of research 
fields~\cite{march,org,optical}, and thus better understanding of the nature 
of projected BCS states is also highly desirable. 
This is precisely the main purpose of this study.

In this paper, one-particle anomalous excitations are studied for 
Gutzwiller-projected BCS superconductors. It is shown that one of the 
characteristic properties for the Gutzwiller-projected BCS states is 
their highly coherent one-particle anomalous excitations. 
It is found that the numerically calculated one-particle anomalous excitations 
can be reproduced quantitatively by a renormalized BCS theory. 
This finding thus strongly indicates that the low-lying excitations of the 
Gutzwiller-projected BCS states are described within a renormalized 
Bogoliubov quasi-particle picture, which resembles the well known 
fact that a Gutzwiller-projected Fermi gas is a Fermi liquid~\cite{voll}. 
The present results also provide a theoretical justification for 
utilizing a simple mean-field based BCS theory to analyze low-energy 
experimental observations for the high-$T_{\rm C}$ cuprate superconductors.

A Gutzwiller-projected BCS state $|\Psi_0^{(N)}\rangle$ with 
$N$ electrons is defined by 
\begin{equation}\label{pbcs}
|{\Psi}_0^{(N)}\rangle={\hat{\cal P}}_N{\hat{\cal P}}_G
|{\rm BCS}\rangle, 
\end{equation}
where $|{\rm BCS}\rangle=\prod_{{\bf k},\sigma} {\hat\gamma}_{{\bf k}\sigma}|0\rangle$ is the ground state of the BCS mean-field 
Hamiltonian~\cite{text} with singlet pairing and 
${\hat\gamma}_{{\bf k}\sigma}$ is 
the standard Bogoliubov quasi-particle annihilation operator with momentum 
${\bf k}$ and spin $\sigma(=\uparrow,\downarrow)$, 
\begin{equation}
\left( 
 \begin{array}{c}
 {\hat\gamma}_{{\bf k}\uparrow}\\
 {\hat\gamma}^\dag_{-{\bf k}\downarrow}
 \end{array}
\right)
=
\left( 
 \begin{array}{cc}
 u_{\bf k}^* & -v_{\bf k}^*\\
 v_{\bf k} & u_{\bf k}
 \end{array}
\right)
\left( 
 \begin{array}{c}
 {\hat c}_{{\bf k}\uparrow}\\
 {\hat c}^\dag_{-{\bf k}\downarrow}
 \end{array}
\right), 
\end{equation} 
${\hat {\cal P}}_G=\prod_{\bf i}(1-{\hat n}_{{\bf i}\uparrow}{\hat n}_{{\bf i}\downarrow})$ is 
the Gutzwiller projection operator excluding sites doubly occupied by 
electrons, and 
${\hat{\cal P}}_N$ the projection operator onto even number $N$ of 
electrons. 
${\hat c}_{{\bf k}\sigma}=\sum_{\bf i} 
{\rm e}^{-i{\bf k}\cdot{\bf i}}{\hat c}_{{\bf i}\sigma}/{\sqrt{L}}$ 
($L$: number of sites) is the 
Fourier transform of the electron annihilation operator 
${\hat c}_{{\bf i}\sigma}$ 
at site ${\bf i}$ with spin $\sigma$, 
and ${\hat n}_{{\bf i}\sigma}={\hat c}_{{\bf i}\sigma}^\dag{\hat c}_{{\bf i}\sigma}$. Note that, since the number $N$ of electrons is even, 
$|\Psi^{(N)}_0\rangle$ is a spin singlet with zero total momentum. 
In the following, it is implicitly assumed that the gap function 
in $|{\rm BCS}\rangle$ is real and the spatial dimensionality is 
two dimensional (2D). However, the generalization of the present study is 
straightforward.

A single-hole (single-electron) added excited state 
$|\Psi^{(N-1)}_{{\bf k}\sigma}\rangle$ 
($|\Psi^{(N+1)}_{{\bf k}\sigma}\rangle$) 
is similarly  
defined~\cite{zhang,rand,ong,yunoki,yunoki2} by 
\begin{equation}\label{pbcs2}
|\Psi^{(N\pm1)}_{{\bf k}\sigma}\rangle={\hat{\cal P}}_{N\pm1}{\hat{\cal P}}_G
{\hat\gamma}_{{\bf k}\sigma}^\dagger |{\rm BCS}\rangle, 
\end{equation}
which has momentum ${\bf k}$, 
total spin $1/2$, and $z$-component of total spin $\sigma$. 
Hereafter, the normalized wave functions for the $N$- and 
$(N\pm1)$-particle states are denoted simply by $|\psi^{(N)}_0\rangle$ and 
$|\psi^{(N\pm1)}_{{\bf k}\sigma}\rangle$, respectively. 
The quasi-particle weights for the one-particle added and removed normal 
excitations are thus defined as 
\begin{equation}
Z_{+}^{(N)}({\bf k}\sigma)=\left|\langle\psi^{(N+1)}_{{\bf k}\sigma}|
{\hat c}^\dag_{{\bf k}\sigma}|\psi^{(N)}_0\rangle\right|^2 \label{z+}
\end{equation}
and
\begin{equation}
Z_{-}^{(N)}({\bf k}\sigma)=\left|\langle\psi^{(N-1)}_{-{\bf k}{\bar \sigma}}|
{\hat c}_{{\bf k} \sigma}|\psi^{(N)}_0\rangle\right|^2, \label{z-}
\end{equation}
respectively, where ${\bar\sigma}$ is the opposite spin to 
$\sigma$.

The one-particle anomalous excitation spectrum is generally defined as
\begin{eqnarray}
&&F({\bf k},\omega)=\nonumber\\
&&-{\frac{1}{\pi}}{\rm Im}\left\langle\Phi_0^{(N+2)}\left|
{\hat c}^\dag_{{\bf k}\uparrow}{\frac{1}{\omega-{\hat H}+{\cal E}_0^{(N)}+i0^+}}
{\hat c}^\dag_{-{\bf k}\downarrow}\right|\Phi_0^{(N)}\right\rangle,\nonumber
\end{eqnarray}
where $|\Phi_n^{(N)}\rangle$ is the $n$th eigenstate ($n=0,1,2,\cdots$, with 
0 corresponding to the ground state) of a system described by Hamiltonian 
${\hat H}$ with its eigenvalue ${\cal E}_n^{(N)}$ and $N$ 
electrons~\cite{ohta}. The spectral representation thus leads
\begin{eqnarray}
&&F({\bf k},\omega)=\sum_{n=0} 
\left\langle\Phi_0^{(N+2)}\left|{\hat c}^\dag_{{\bf k}\uparrow}\right|\Phi_n^{(N+1)}\right\rangle\quad\quad\nonumber\\
&&\ \times\left\langle\Phi_n^{(N+1)}\left|{\hat c}^\dag_{-{\bf k}\downarrow}\right|\Phi_0^{(N)}\right\rangle
\delta\left(\omega-{\cal E}_n^{(N+1)}+{\cal E}_0^{(N)}\right). 
\end{eqnarray} 
Note that $F({\bf k},\omega)$ is a real function provided that spin rotation 
and reflection invariance as well as time reversal symmetry are 
assumed~\cite{fermi}.  
The frequency integral of the one-particle anomalous excitation spectrum, 
$F_{\bf k}^{(N)}=\int_{-\infty}^{\infty}d\omega F({\bf k},\omega)
=\left\langle\Phi_0^{(N+2)}\left|{\hat c}^\dag_{{\bf k}\uparrow}
{\hat c}^\dag_{-{\bf k}\downarrow}\right|\Phi_0^{(N)}\right\rangle$,  
provides a well-known sum rule which will be used later.

To study the one-particle anomalous excitations for the projected BCS state, 
let us first defined the following quantity similar to $F_{\bf k}^{(N)}$ 
for the projected BCS states: 
\begin{equation}\label{z2_def}
Z_2^{(N)}({\bf k})=\left\langle\psi_0^{(N+2)}\left|{\hat c}^\dag_{{\bf k}\uparrow}
{\hat c}^\dag_{-{\bf k}\downarrow}\right|\psi_0^{(N)}\right\rangle.
\end{equation}
Here $|\psi_0^{(N+2)}\rangle$ is constructed in exactly the same way 
as $|\psi_0^{(N)}\rangle$ except that the number of electrons onto which 
the state is projected is $N+2$. Using the previously derived relations 
for the projected BCS states~\cite{yunoki2}, one can now easily show that 
\begin{equation}\label{relation2}
Z_2^{(N)}({\bf k}) =
\left\langle\psi_0^{(N+2)}\left|{\hat c}^\dag_{{\bf k}\uparrow}\right|\psi_{-{\bf k}\downarrow}^{(N+1)}\right\rangle
\left\langle\psi_{-{\bf k}\downarrow}^{(N+1)}\left|{\hat c}^\dag_{-{\bf k}\downarrow}\right|\psi_0^{(N)}\right\rangle, 
\end{equation}
i.e., as will be shown later, the quasi-particle weight for the one-particle 
anomalous excitations is $Z_2^{(N)}({\bf k})$. It is also interesting to 
notice that the above equation relates $Z_2^{(N)}({\bf k})$ to the 
quasi-particle weights for the one-particle normal excitations, i.e., 
\begin{eqnarray}\label{relation}
\left|Z_2^{(N)}({\bf k})\right|^2
&=&Z_+^{(N)}(-{\bf k}\downarrow)\cdot Z_-^{(N+2)}({\bf k}\uparrow),  
\end{eqnarray} 
which should be useful for computing $Z_-^{(N)}({\bf k}\sigma)$~\cite{note}.

\begin{figure}[hbt]
\includegraphics[clip=true,width=7.5cm,angle=-0]{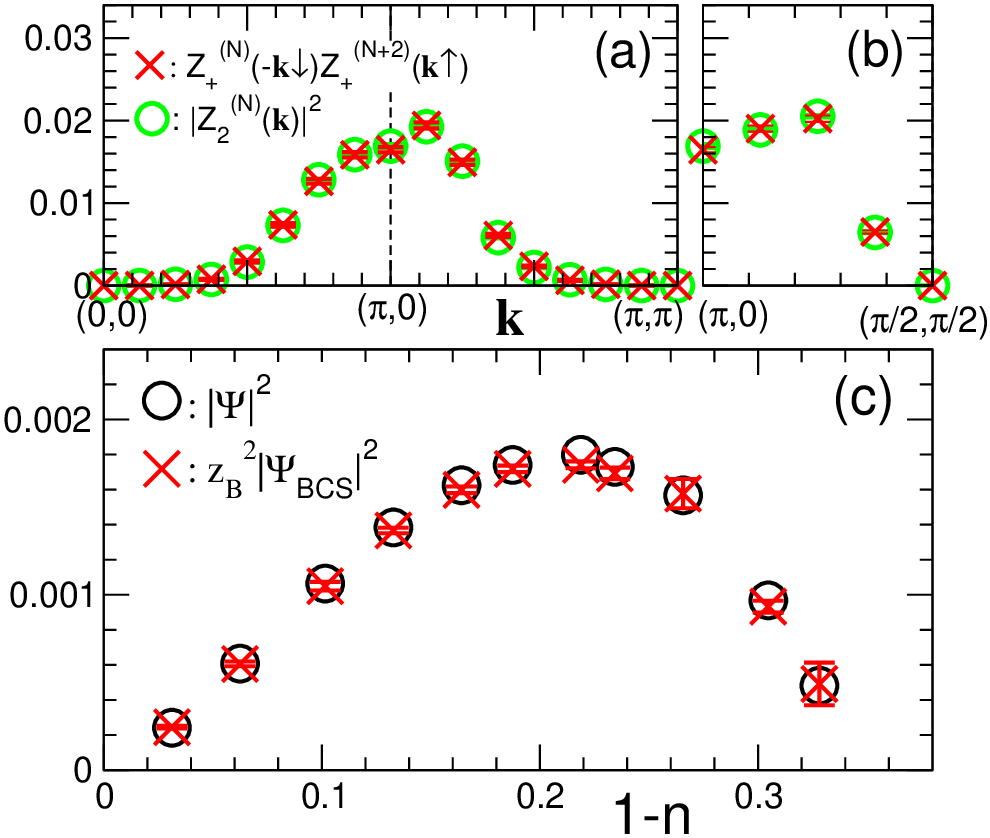}
\begin{center}
\caption{
(Color online) 
(a) and (b): $\left|Z_2^{(N)}({\bf k})\right|^2$ (circles) and 
$Z_+^{(N)}(-{\bf k}\downarrow)\cdot Z_-^{(N+2)}({\bf k}\uparrow)$ (crosses) 
for different momenta ${\bf k}$ and $N=222$ $(n\approx0.87)$. 
(c): $|\Psi|^2=|\langle\psi_0^{(N+2)}|c_{{\bf i}\uparrow}^\dag c_{{\bf i}+{\bf x}\downarrow}^\dag|\psi_0^{(N)}\rangle|^2$ (circles) as a function of 
electron density $n$. $|\psi_0^{(N)}\rangle$ is optimized for 
the 2D $t$-$t'$-$J$ model with $t'/t=-0.2$ and $J/t=0.3$ on an 
$L=16\times16$ cluster with periodic boundary conditions~\cite{yunoki,note3}. 
For comparison, $z_B^2|\Psi_{\rm BCS}|^2$ for a renormalized BCS theory 
(see the text) is also presented by crosses in (c). 
}
\label{z2}
\end{center}
\end{figure}

The validity of Eq.~(\ref{relation}) can be checked numerically by 
computing all quantities, $Z_+^{(N)}(-{\bf k}\downarrow)$, 
$Z_-^{(N+2)}({\bf k}\uparrow)$, and $\left|Z_2^{(N)}({\bf k})\right|^2$, 
independently.  
A typical set of results calculated by a standard variational Monte Carlo 
technique on finite clusters is shown in Figs~\ref{z2} (a) and (b), 
where one can see that indeed Eq.~(\ref{relation}) holds within the 
statistical error~\cite{note2}.

Now let us assume that there exists a system (with Hamiltonian 
${\hat{H}}$) 
for which the ground state 
and the low-energy excited states can be described approximately by the 
projected BCS states $|\Psi_0^{(N)}\rangle$ and 
$|\Psi^{(N\pm1)}_{\bf k}\rangle$, i.e., 
$|\Phi_0^{(N)}\rangle\approx|\Psi_0^{(N)}\rangle$ and 
$|\Phi_n^{(N+1)}\rangle\approx|\Psi^{(N+1)}_{\bf k}\rangle$, etc. 
One immediate consequence of this assumption is that the sum rule for 
$F({\bf k},\omega)$ is now 
$\int_{-\infty}^{\infty}d\omega F({\bf k},\omega) = Z_2^{(N)}({\bf k})$. 
Another consequence is that the spectral representation of the one-particle 
anomalous excitations is 
\begin{eqnarray}\label{spectrum}
F({\bf k},\omega)&=&Z_2^{(N)}({\bf k})\cdot\delta\left(\omega-E_{-{\bf k}}^{(N+1)}+E_0^{(N)}\right)\nonumber\\
&+&\sum_{n(\ne0)}\,({\rm other\ terms}), 
\end{eqnarray}
i.e., the quasi-particle weight for the one-particle anomalous excitations 
is $Z_2^{(N)}({\bf k})$. Here $E_0^{(N)}= E({\Psi}_0^{(N)})
=\langle{\psi}_0^{(N)}|{\hat H}|{\psi}_0^{(N)}\rangle$, and 
$E_{-{\bf k}}^{(N+1)}=E({\Psi}_{-{\bf k}\downarrow}^{(N+1)})$. 
This is because of the equality derived here in Eq.~(\ref{relation2}). 
Since the spectral weight is not positive 
definite, the above equation along with the sum rule 
does not immediately imply that the contribution of incoherent 
``other terms'' in Eq.~(\ref{spectrum}) is negligible. 
However, using the property of the projected BCS states reported 
previously that the one-particle added normal excitations are 
coherent~\cite{ong,mohit,yunoki2}, one can easily show that indeed the 
one-particle anomalous excitations consist of a single 
coherent part for each ${\bf k}$ with no incoherent contributions. 
It is interesting to note that numerically exact diagonalization 
studies of small clusters have also found that the one-particle anomalous
excitations for the 2D $t$-$J$ model are highly coherent with relatively 
small incoherent contributions~\cite{ohta}.

Let us now calculate the one-particle anomalous excitation 
spectrum $F({\bf k},\omega)$ for the projected BCS states. 
For this purpose, here we will consider the 2D $t$-$t'$-$J$ model 
on the square lattice~\cite{yunoki}. 
This model has been studied extensively and found to show a $d$-wave 
superconducting regime in the phase diagram~\cite{tj,ohta}. 
Furthermore, it is well-known that a Gutzwiller-projected BCS state 
with $d$-wave pairing symmetry [Eq.~(\ref{pbcs})] is a faithful 
variational ansatz for the superconducting state of this 
model~\cite{yokoyama}. The model parameters used here are set to be 
$t'/t=-0.2$ and $J/t=0.3$~\cite{elbio}.

\begin{figure}[hbt]
\includegraphics[clip=true,width=8.cm,angle=-0]{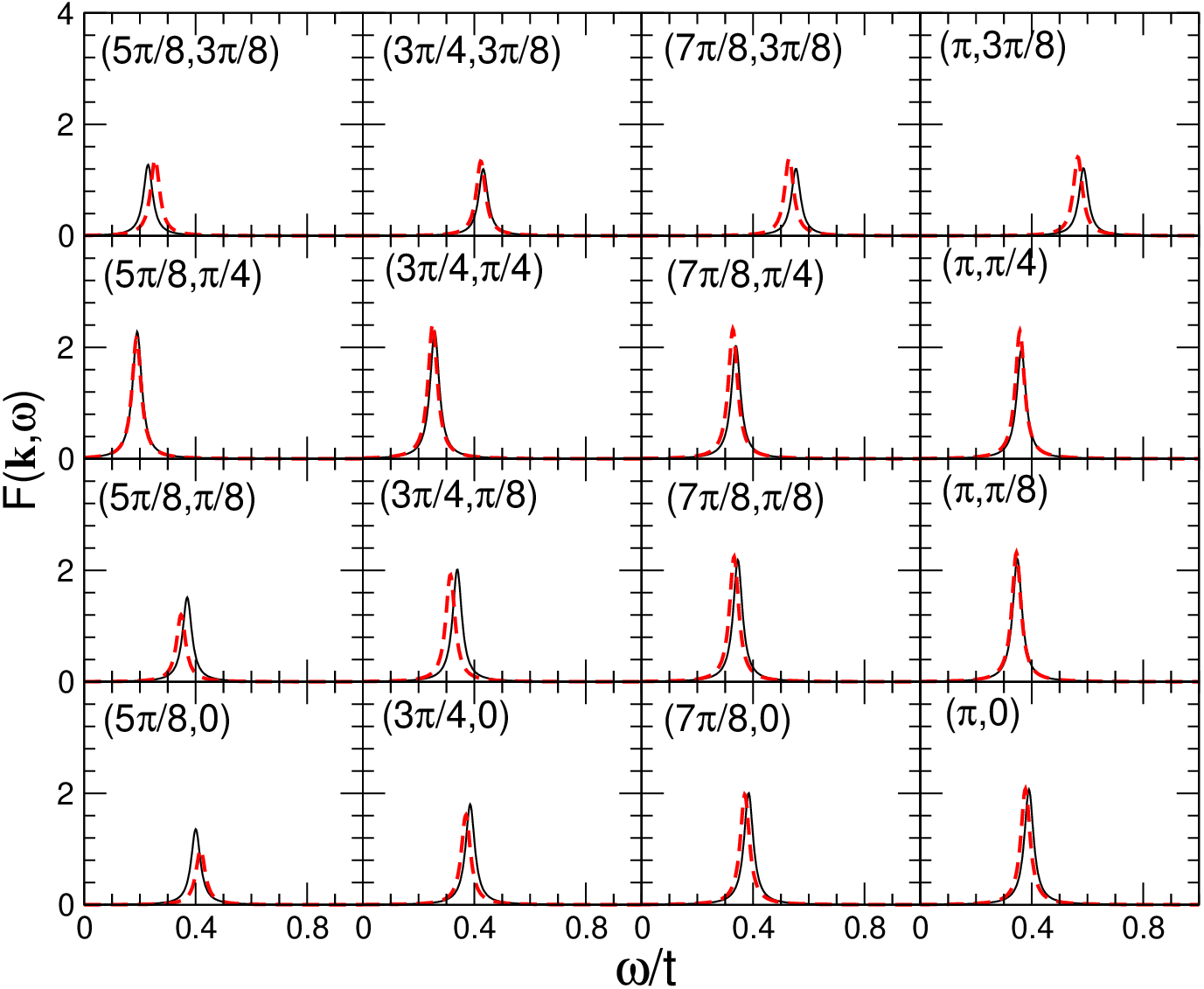}
\begin{center}
\caption{
(Color online) 
The one-particle anomalous excitation spectrum $F({\bf k},\omega)$ for the 
projected BCS states [Eq.~(\ref{spectrum})] (solid lines). 
$|\Psi_0^{(N)}\rangle$ used here is optimized for the 2D $t$-$t'$-$J$ model 
with $t'/t=-0.2$, $J/t=0.3$, and $N=222$ on an $L=16\times16$ cluster 
$(n\approx0.87)$ with periodic boundary conditions~\cite{note3}. 
The momenta ${\bf k}$ studied are indicated in the figures. 
For comparison,  $F({\bf k},\omega)$ for a renormalized BCS theory 
(see the text) is also presented by dashed lines. The momentum-independent 
renormalization factor $z_{B}$ is 0.30. 
The delta function $\delta(\omega)$ is represented by a Lorentzian 
function $\varepsilon/{\pi(\omega^2+\varepsilon^2)}$ with $\varepsilon=0.02t$. 
}
\label{spec}
\end{center}
\end{figure}

The results of the numerically calculated $F({\bf k},\omega)$ for 
representative momenta are shown in Fig.~\ref{spec}, where 
$|\Psi_0^{(N)}\rangle$ is optimized to 
minimize the variational energy for $N=222$ on an $L=16\times16$ cluster 
$(n\approx0.87)$ with periodic boundary conditions~\cite{note3}. As is expected 
for a $d$-wave superconductor, the spectral weight becomes smaller toward the 
nodal line in the $(0,0)$-$(\pi,\pi)$ direction 
[see also Figs.~\ref{z2} (a) and (b)], 
and it changes the sign across the nodal line where the weight is zero.

\begin{figure}[hbt]
\includegraphics[clip=true,width=8.cm,angle=-0]{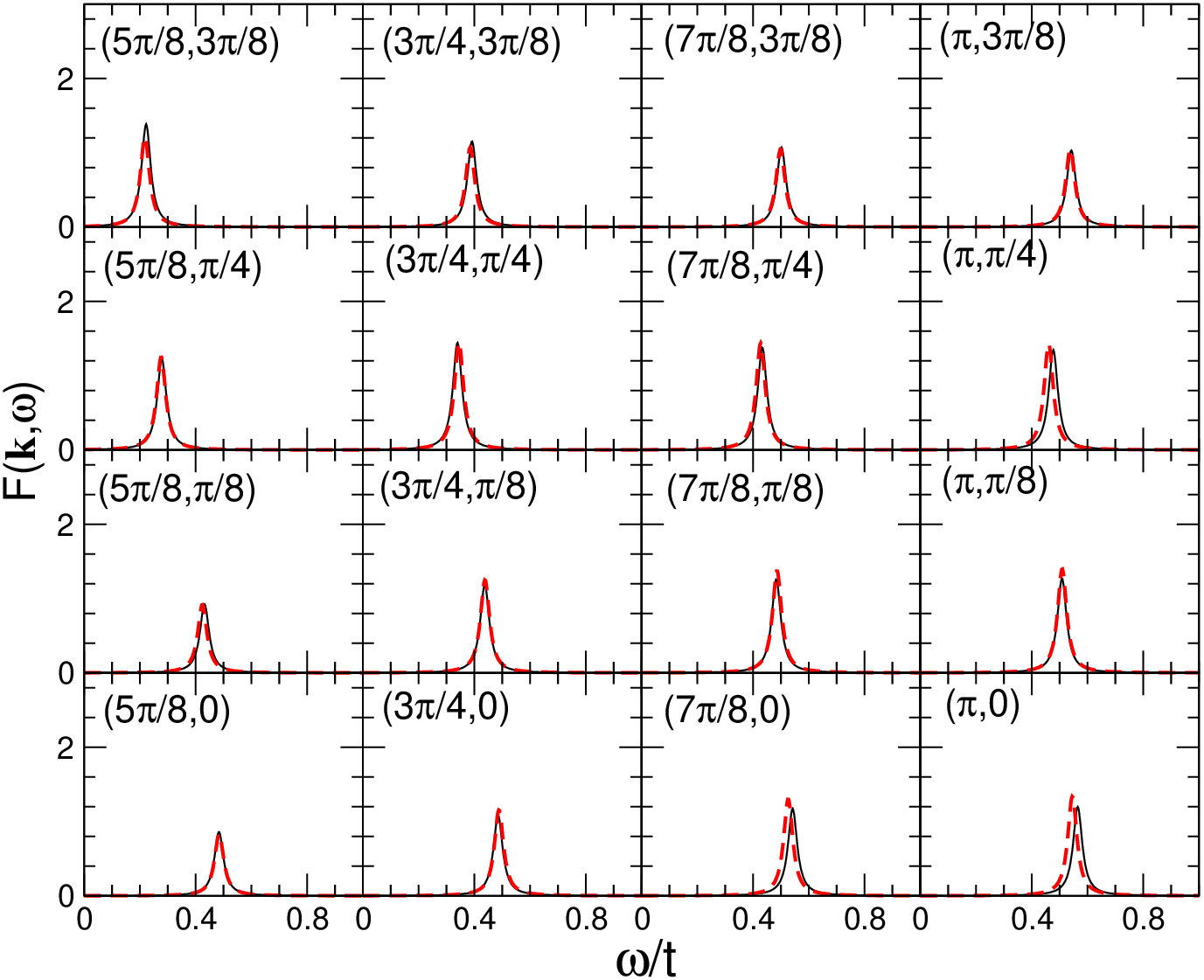}
\begin{center}
\caption{
(Color online) 
The same as in Fig.~\ref{spec} but for $N=240$ $(n\approx0.94)$~\cite{note3}. 
The momentum-independent renormalization factor $z_{B}$ is 0.18.
}
\label{spec2}
\end{center}
\end{figure}

To understand the nature of the low-lying excitations observed in 
$F({\bf k},\omega)$, here the results are analyzed based on a renormalized 
BCS theory with $d$-wave pairing symmetry~\cite{text}. 
The procedure adopted is as follows: (i) the excitation energy 
$E({\bf k})=E_{\bf k}^{(N+1)}-E_0^{(N)}$ is fitted for all momenta 
${\bf k}$ in the 
whole Brillouin zone by a standard Bogoliubov excitation 
spectrum~\cite{note6}, (ii)  using the fitting parameters determined in (i), 
the BCS spectral weight for the one-particle anomalous excitations, 
${u^{\rm (BCS)}_{\bf k}}^* v^{\rm (BCS)}_{\bf k}$, is 
calculated, and (iii) the BCS spectrum is renormalized by a 
momentum independent constant $z_B$ in such a way that 
\begin{equation}\label{zb}
\sum_{\bf k}\left|Z_2^{(N)}({\bf k})\right|^2=z_B^2
\sum_{\bf k}\left|u^{\rm (BCS)}_{\bf k}\cdot v^{\rm (BCS)}_{\bf k}\right|^2.
\end{equation} 
The obtained renormalized BCS spectra are shown in Fig.~\ref{spec} by 
dashed lines. It is 
clearly seen in Fig.~\ref{spec} that the renormalized BCS spectra can 
reproduce almost quantitatively $F({\bf k},\omega)$ for the projected BCS 
states. It should be emphasized that the procedure employed above is highly 
nontrivial and it is beyond a simple fitting of numerical data. 
Similar agreement is also found for different sets of 
model parameters, one of which is exemplified in Fig.~\ref{spec2}. 
The surprisingly excellent agreement found here strongly indicates that the 
low-lying excitations described by the projected BCS states can be well 
understood within a renormalized Bogoliubov quasi-particle picture.

To further examine the validity of the renormalized Bogoliubov 
quasi-particle picture for the projected BCS states, let us finally study 
the superconducting order parameter, which is here defined as 
$\Psi=\langle\psi_0^{(N+2)}|c_{{\bf i}\uparrow}^\dag c_{{\bf i}+{\bf x}\downarrow}^\dag|\psi_0^{(N)}\rangle$ (${\bf x}$ 
being the unit vector in the $x$ direction). 
The electron density ($n$) dependence of $\Psi$ is shown in Fig.~\ref{z2} (c) 
for the 2D $t$-$t'$-$J$ model,  
where $|\psi_0^{(N)}\rangle$ is optimized for each $n$~\cite{note3}. 
As seen in Fig.~\ref{z2} (c), $|\Psi|^2$ vs $n$ shows a domelike 
behavior, similar to the pairing correlation function at the maximum 
distance as a function of $n$ reported before~\cite{rand}. 
It is also interesting to notice 
that $|\Psi|^2$ is proportional to $1-n$ for small $1-n$. 
The corresponding quantity $\Psi_{\rm BCS}$ for the BCS state with 
${u^{\rm (BCS)}_{\bf k}}$ and $v^{\rm (BCS)}_{\bf k}$, determined by the 
procedure mentioned above, can also be calculated by 
$\Psi_{\rm BCS}={\frac{1}{L}}\sum_{\bf k}{\rm e}^{i{\bf k}\cdot{\bf x}}{u^{\rm (BCS)}_{\bf k}}^* v^{\rm (BCS)}_{\bf k}$. 
If a renormalized Bogoliubov quasi-particle picture is valid, 
$\Psi \cong z_B \Psi_{\rm BCS}$ is expected. As seen in Fig.~\ref{z2} (c), 
this is in fact clearly the case. This result also gives a clear 
physical meaning to the renormalization factor $z_B$ introduced in 
Eq.~(\ref{zb}).

As is well known, a Gutzwiller-projected Fermi gas is described within a 
Fermi liquid picture~\cite{voll}. The present results thus strongly 
suggest that analogously a Gutzwiller-projected, correlated BCS state 
(``projected BCS gas'') can still be described within a renormalized 
BCS-Bogoliubov quasi-particle picture (``BCS liquid'')~\cite{note4}. 
This is in fact in accordance with recent photoemission spectroscopy 
experiments on high-$T_{\rm C}$ cuprate superconductors for which 
low-lying excitations consistent with a BCS theory have been 
revealed~\cite{takahashi}. Moreover, the present results would also 
provide a theoretical justification for 
employing a mean-field-based BCS-like theory to analyze the low-energy 
dynamics observed experimentally in the superconducting state of 
high-$T_{\rm C}$ cuprate superconductors~\cite{bala}.

To summarize, 
the one-particle anomalous excitations have been studied 
to understand the nature of the low-lying excitations of strongly correlated 
superconductors described by the Gutzwiller-projected BCS states. 
It was found that the low-lying excitations, which are highly coherent, can 
be essentially described within a renormalized Bogoliubov quasi-particle 
picture. This finding thus resembles the well-known result that a 
Gutzwiller-projected Fermi gas is a Fermi liquid. Finally, the present 
study has demonstrated that a variational Monte Carlo-based approach can be 
also utilized to explore low-lying excitations, 
and hopefully this work will stimulate further 
studies in this direction for other dynamical quantities.

The author would like to acknowledge useful discussion with S. Sorella, 
T. K. Lee, D. Ivanov, E. Dagotto, C.-P. Chou, S. Bieri, R. Hlubina, and 
S. Maekawa. This work was supported in part by INFM.


\end{document}